\documentclass[aps,pre,twocolumn,showpacs,superscriptaddress]{revtex4-1}
\usepackage[unicode]{hyperref}
\usepackage{amsmath}
\usepackage[usenames]{color}
\usepackage{color}
\usepackage{colortbl}
\definecolor{linkcolor}{rgb}{0.8,0,0.2}
\definecolor{citecolor}{rgb}{0,0.6,0.2}
\definecolor{urlcolor}{rgb}{0,0,1}
\hypersetup{
    colorlinks, linkcolor={linkcolor},
    citecolor={citecolor}, urlcolor={urlcolor}}
\usepackage{graphicx}
\usepackage{epstopdf}
\usepackage[utf8]{inputenc}
\usepackage[english]{babel}

\begin{document}
\title{Model of the Tamm plasmon polariton based solar cell}
\author{Rashid G.~Bikbaev}
\email{bikbaev@iph.krasn.ru}
\affiliation{L.~V.~Kirensky Institute of Physics, Federal Research Center KSC SB RAS, Krasnoyarsk, 660036, Russia}
\affiliation{Siberian Federal University, Krasnoyarsk, 660041, Russia}
\author{Dmitrii Pykhtin}
\affiliation{Siberian Federal University, Krasnoyarsk, 660041, Russia}
\author{Stepan Ya. Vetrov}
\affiliation{Siberian Federal University, Krasnoyarsk, 660041, Russia}
\affiliation{L.~V.~Kirensky Institute of Physics, Federal Research Center KSC SB RAS, Krasnoyarsk, 660036, Russia}
\author{Ivan V.~Timofeev}
\affiliation{L.~V.~Kirensky Institute of Physics, Federal Research Center KSC SB RAS, Krasnoyarsk, 660036, Russia}
\affiliation{Siberian Federal University, Krasnoyarsk, 660041, Russia}
\author{Vasiliy F.~Shabanov}
\affiliation{L.~V.~Kirensky Institute of Physics, Federal Research Center KSC SB RAS, Krasnoyarsk, 660036, Russia}

\begin{abstract}

A model of an organic solar cell based on a Tamm plasmon polariton localized at the interface between a photonic crystal and a photosensitive layer with an embedded square or hexagonal plasmonic array has been proposed. The spectral properties of the structures have been investigated in the framework of the temporal coupled mode theory and confirmed by the transfer matrix method. It has been shown that the conjugation of a photonic crystal with a photosensitive layer leads to the excitation of a Tamm plasmon polariton at the interface between them and, consequently, to an increase in the integral absorption by 12\% as compared with a planar solar cell without a photonic crystal. 
\end{abstract}

\maketitle

\section{Introduction}
The progress in solar energy goes towards enhancing the energy conversion efficiency with simultaneous increasing the reliability and decreasing the cost of solar cells. Accordingly, in the last fifteen years, the conjugated polymer-based organic solar cells (OSCs) have been rapidly developed, which are attractive by their low cost and weight and the mechanical flexibility of solar panels ~\cite{Atwater2010PlasmonicsDevices,Heeger201425thOperation,He2015Single-junctionPhotovoltage}. 
Since such OSCs contain a bulk heterojunction, it is necessary to find a compromise between the photon absorption and carrier transport efficiencies.
In this case, the thickness of a photosensitive layer (PSL) is no larger than 100~nm, which significantly limits the efficiency of absorption of incident light.
In view of the aforesaid, methods for manipulating light for increasing the absorption in the PSL by means of internal scattering or the plasmon resonance effect have found wide applications.
In particular, the authors of ~\cite{Duche2009ImprovingContribution} experimentally demonstrated the possibility of increasing the absorption of light in a photoactive layer containing silver nanoparticles.
In ~\cite{Li2013EfficiencyNanoparticles}, it was shown that the absorption line of the PSL can be broadened by embedding particles of various shapes into it.
This direction has gained wide use.
The PSL was added with nanowires~\cite{Kim2011SilverApplications}, nanorods~\cite{He2015Single-junctionPhotovoltage}, and particles in the form of cubes, dodecahedra, octahedra, and triangular plates~\cite{Tseng2015Shape-DependentEffect}.

Another important way of increasing the integral absorption in the PSL is the introduction of a photonic crystal (PhC) comprising WO$_3$/LiF~\cite{Yu2012SemitransparentCrystals,Yu2013SimultaneousCrystals,Yu2014LightReflectors} and TiO$_2$/SiO$_2$~\cite{Lunt2011TransparentApplications} layers into a 1D OSC. 
The high reflectivity in the OSC band gap ensures the repeated transmission of the light incident onto the structure through the PSL, thereby increasing the OSC efficiency.
Recently, we have proposed to use the PSL doped with plasmonic nanoparticles as a mirror confining a 1D PhC ~\cite{Bikbaev2021}. 
In this case, a Tamm plasmon polariton (TPP) is localized at the PSL/PhC interface ~\cite{Kaliteevski2007,Sasin2008,Vetrov2013,Vetrov2017,Bikbaev2019Epsilon-Near-ZeroPolariton}, which leads to the occurrence of an additional absorption band of the radiation incident onto the structure and, consequently, increases the efficiency of the OSC. This structure makes it possible to entirely exclude a metallic contact and, thus, ensure the absorption mainly in the PSL.
The dispersion of the PSL doped with plasmon nanoparticles was, in this case, determined using the effective medium model.
In this study, we examined not chaotically dispersed nanoparticles, but 2D square and hexagonal arrays and investigated their optical properties by the finite difference time domain (FDTD) method. Importantly, such structures can be obtained by self-assembly, which significantly reduces their cost and speeds up the production~\cite{Huh2020}.

\section{Description of the Model}

\begin{figure}[h]
    \centering
    \includegraphics[scale=0.5]{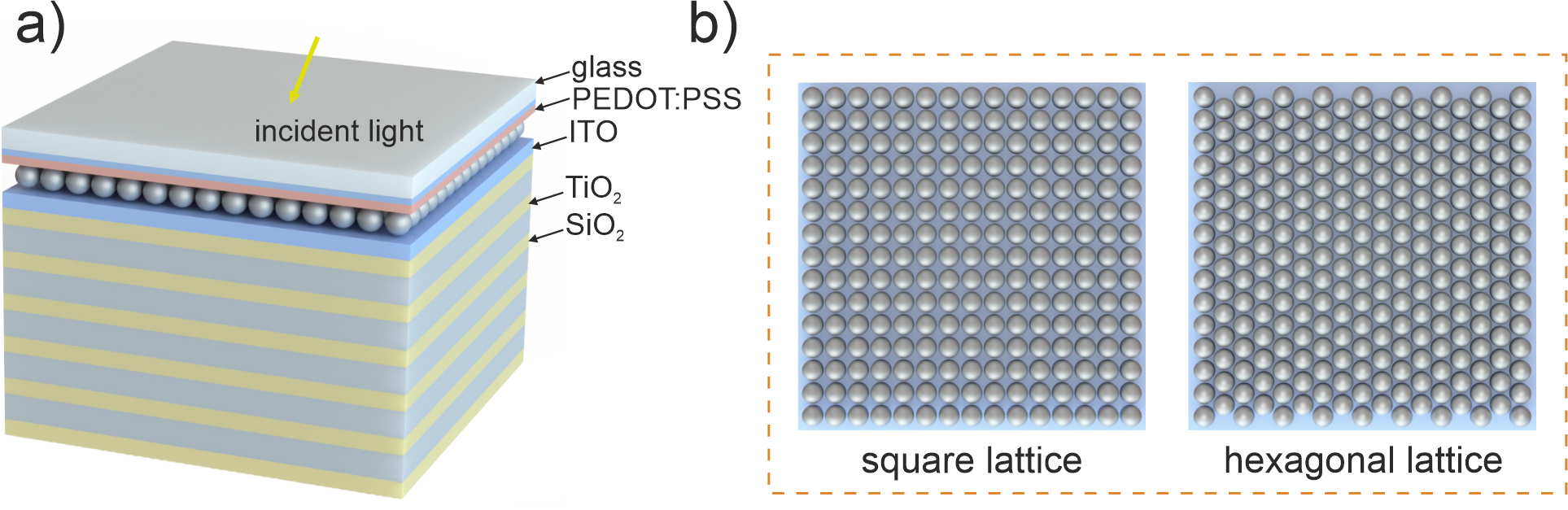}
    \caption{(a) Schematic of the investigated structure and (b) two types of arrays.}
    \label{fig:fig1}
\end{figure}

A schematic of the investigated solar cell is shown in Fig. ~\ref{fig:fig1}a. 
The P3HT:PC61BM PSL ~\cite{Stelling2017PlasmonicCells} with a thickness of 70~nm is doped with silver nanospheres with a radius of $r = 30$~nm. The distance between the sphere centers is $p=65$~nm. In this case the filling factor $f$, i.e., the volume fraction of nanoparticles in the PSL equal to 40\%. The permittivity of silver was taken from~\cite{lide1995crc}. The thickness of the poly(3,4-ethylenedioxythiophene) polystyrene sulfonate (PEDOT:PSS) layer~\cite{Chen2015OpticalCells} is 20~nm. 
The contacts used were ITO films with thicknesses of 15~ and 45~nm. 

A PhC unit cell was formed from silicon dioxide SiO$_2$~\cite{Malitson1965InterspecimenSilica} and titanium dioxide TiO$_2$~\cite{DeVore1951RefractiveSphalerite} with thicknesses of $d_{SiO_2} = 45$~nm and $d_{TiO_2} = 40$~nm respectively. 

The effective refractive index of the square and hexagonal array (see Fig.~\ref{fig:fig1}b) embedded in the PSL  was obtained by numerical
calculations (s-parameter retrieval method~\cite{Smith2005}).
The effective refractive index of the PSL, in this case, can be defined as:

\begin{equation}
  n_{\mathrm{eff}}=\frac{1}{kd}\ {cos}^{-1}\left(\frac{1-S^2_{11}+S^2_{21}}{2S_{21}}\right),
\end{equation}

\noindent where $d$ is the PSL thickness, $k$ is the wave vector, and $S_{11}$ and $S_{21}$ are the components of the scattering matrix.

\begin{figure}
    \centering
    \includegraphics[scale=0.5]{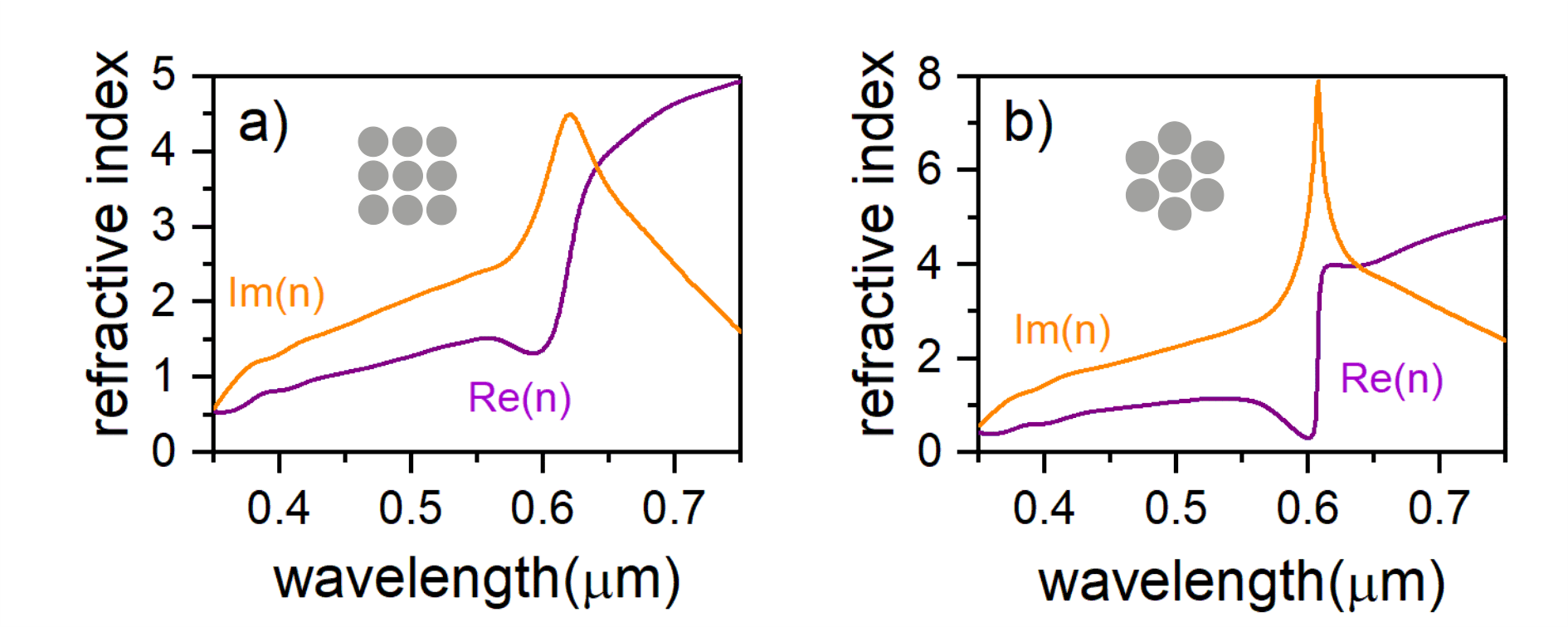}
     \caption{Real and imaginary parts of the refractive index for (a) the square and (b) hexagonal arrays of silver nanospheres determined using the s-parameters.}
    \label{fig:fig2}
\end{figure}

The dependences of the real and imaginary parts of the effective refractive index of the PSL with the square and hexagonal arrays on the incident radiation wavelength are presented in Fig.~\ref{fig:fig2}.
For both the square and hexagonal array, the real part of the refractive index is maximum at a wavelength of 600 nm, which is caused by the plasmon resonance in nanoparticles.

\section{Results and Discussion}
\subsection{Coupled mode theory}

According to temporal coupled mode theory~\cite{haus1984waves,joannopoulos2008photonic}, any state (resonance) has its own frequency $\omega_0$ and number $N$ of energy channels outside and inside the resonance.
In this case, the energy loss in the channels is described by the relaxation times $\tau_l$ or relaxation rate $\gamma_l=1/\tau_l$, where $l=1...N$. 
If the energy leaves the state along two energy channels with relaxation times  $\tau_1$ and $\tau_2$, then the
relaxation time of the state is determined as $1/\tau = 1/\tau_1 + 1/\tau_2$. 
In the presented solar cell three energy channels contribute to the TPP formation.
We denote the energy relaxation to the PSL transmission, absorption and PhC transmission channel as $\gamma_{PSL}$ , $\gamma_{A}$ and $\gamma_{PhC}$, respectively. 
Since the energy accumulated in the TPP is the same for determining the rate of relaxation to each channel, the relaxation rates and corresponding energy coefficients of the structure are related as  \cite{Yang2017}:

 \begin{equation}
 \gamma _{PSL} :\gamma _{A} :\gamma _{PhC} =T_{PSL} :A_{PSL} :T_{PhC}.
 \label{eq:Eq10}
 \end{equation}

In the case of opaque PhC, its relaxation channel can be ignored. As a result, critical coupling condition~(\ref{eq:Eq10}) can be written in the form:
\noindent 
\begin{equation}
\gamma _{PSL} =\gamma _{A} ;\quad\gamma _{PhC} =0\Leftrightarrow T_{PSL} =A_{PSL} ;\quad T_{PhC} =0.
\label{eq:Eq11}
\end{equation}

This equation can be solved graphically. 
To do this, we should build the dependence $A_{PSL}(T_{PSL})$. 
We consider the PSL film with refractive index $n_2$, which is located between two dielectric media with the refractive indices $n_1=n_3=1$. The transmittance, reflectance and absorptance
of the PSL film are determined using the Airy formulas

\noindent
\begin{equation}
\begin{gathered}
T_{PSL}=\frac{n_3}{n_1}\bigg|\frac{t_{12}+t_{23}e^{i\beta}}{1+r_{12}r_{23}e^{2i\beta}}\bigg|^2,\quad
R_{PSL}=\bigg|\frac{r_{12}+r_{23}e^{2i\beta}}{1+r_{12}r_{23}e^{2i\beta}}\bigg|^2,\\
A_{PSL}=1-T_{PSL}-R_{PSL},
\label{eq:Eq13}
\end{gathered}
\end{equation}
where $\beta={2\pi}n_2d_\textit{PSL}/{\lambda_0}$ is the phase incoming during the passage of the layer by the wave; $\lambda_0$ is
the wavelength; $d_\textit{PSL}$ is the TCO film thickness;  $t_{12}=2n_1/(n_1 + n_2)$, $r_{12}=(n_1-n_2)/(n_1+n_2)$ and $t_{23}=2n_2/(n_2 + n_3)$, $r_{23}=(n_2-n_3)/(n_2+n_3)$ --  are the amplitudes of transmission and reflection at the interfaces 1-2 and 2-3.

\begin{figure}[h!]
\centering
\includegraphics[scale=0.6]{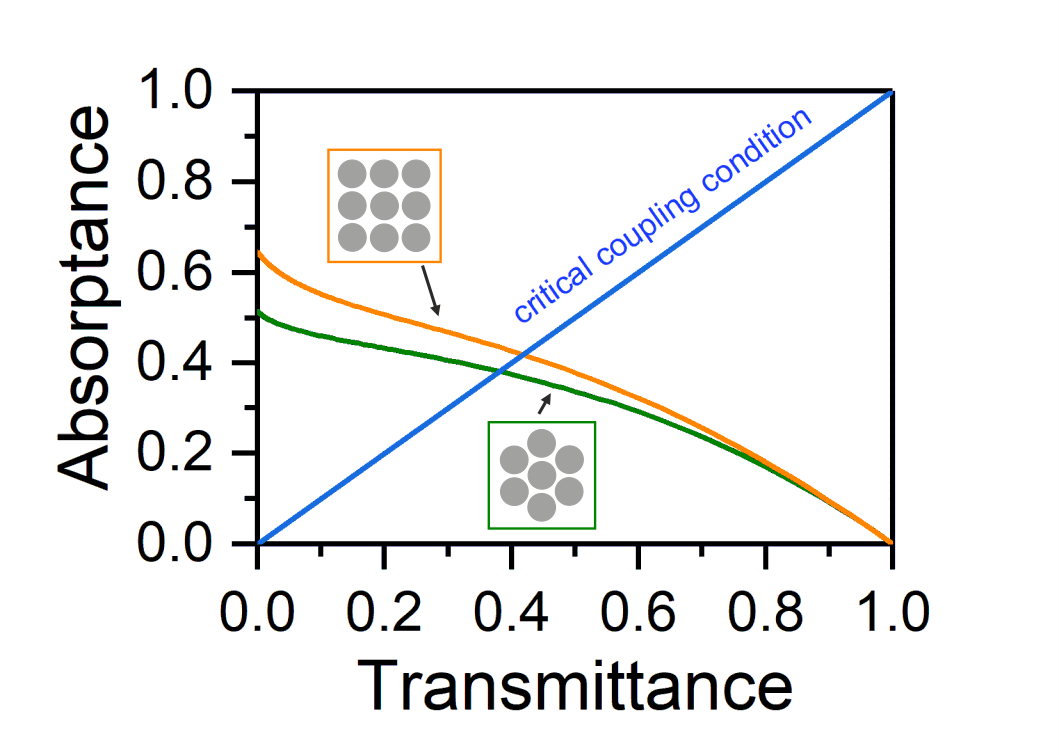}
\caption{Dependence of the absorptance of the PSL films on their transmittance at different film thicknesses.
The points of intersection of the lines with the blue line meet critical coupling conditions~(Eq.~\ref{eq:Eq11}).}
\label{fig:Fig3}
\end{figure}

The obtained results are presented in Fig.~\ref{fig:Fig3}.
It can be seen from this figure that for PSL film with hexagonal array the critical coupling
condition is established at the lower transmittance and absorptance compared to square array. In this comparison we assume that the resonance spectral linewidth is determined by the total energy relaxation rate. 
Thus, the resonance line and, consequently, the absorption band of the PSL film with hexagonal array are narrower than for the PSL film with square array film. In addition, it means that in the case of a PhC bounded by PSL film with hexagonal array, the TPP Q factor is larger.

\subsection{Transfer matrix simulation}

To verified this fact, we calculated the integral absorption of the PSL layer by the transfer matrix method~\cite{Yeh1979}.
The integral absorption is meant to be the PSL absorption normalized to the solar radiation spectrum.
Generally, the integral absorption is determined for each polarization, $A_{TE}$ and $A_{TM}$, separately and their arithmetic mean $A_{total}=(A_{TE}+A_{TM})/2$ yields the total absorption in the layer. 
At the normal incidence, we have $A_{total}= A_{TE} = A_{TM}$, which can be determined as:

\begin{equation}
A_{total} = \frac{\int\limits^{\lambda_2}_{\lambda_1} A(\lambda)S(\lambda)\,d\lambda}{\int\limits^{\lambda_2}_{\lambda_1} S(\lambda)\,d\lambda},  
\end{equation}

\noindent where $\lambda_1=350$~nm, $\lambda_2=500$~nm, $A(\lambda)$ is the absorption in the PSL, and $S(\lambda)$ is the solar radiation spectrum (AM1.5).

\begin{figure}
    \centering
    \includegraphics[scale=0.5]{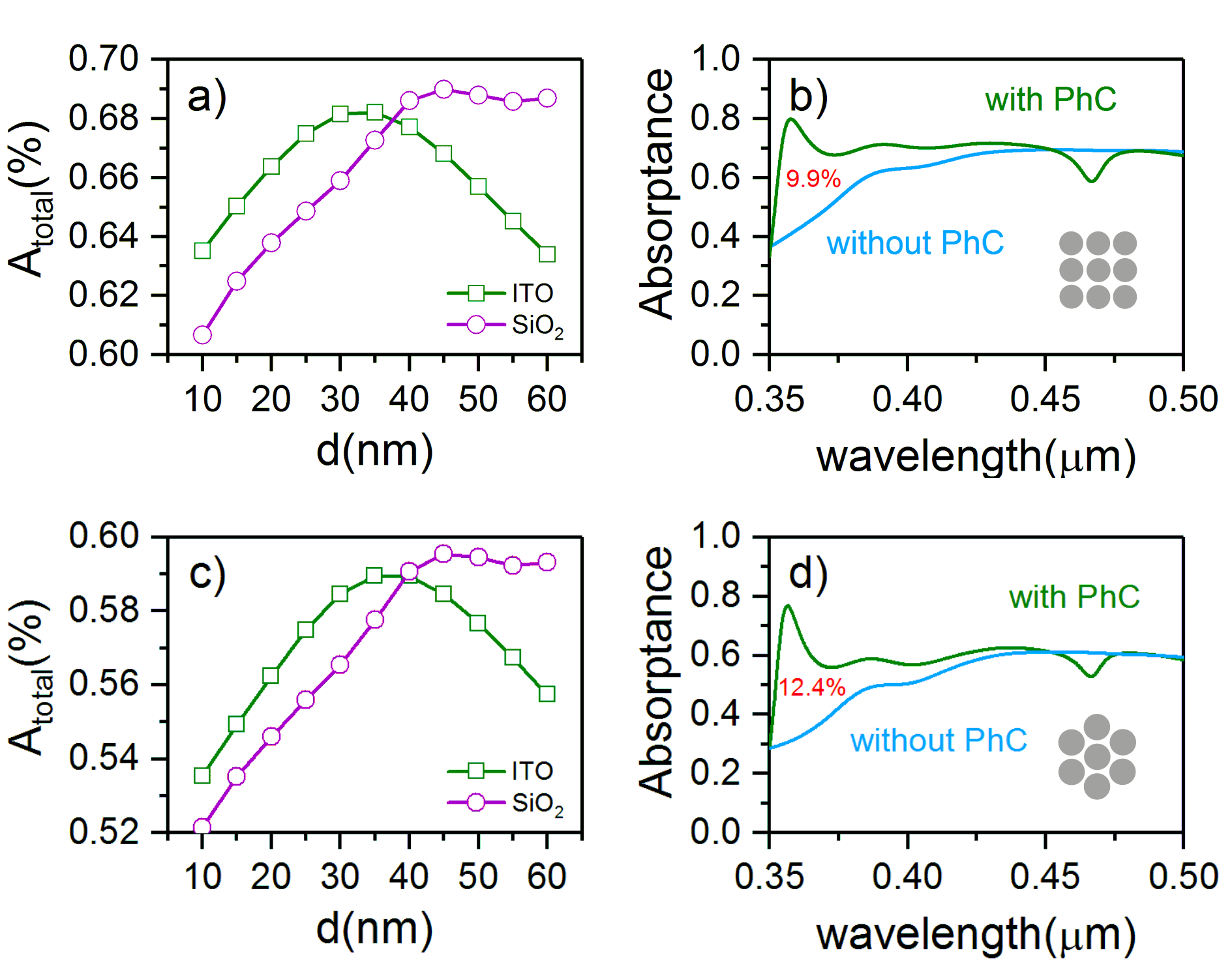}
      \caption{ Integral absorption in the PSL versus ITO and $SiO_2$ film thickness for (a) the square and (c) hexagonal arrays. Absorption in the PSL for the structures with and without PhC in the case of (b) the square and (d) hexagonal array.}
    \label{fig:fig3}
\end{figure}

\begin{figure}[h]
    \centering
    \includegraphics[scale=0.5]{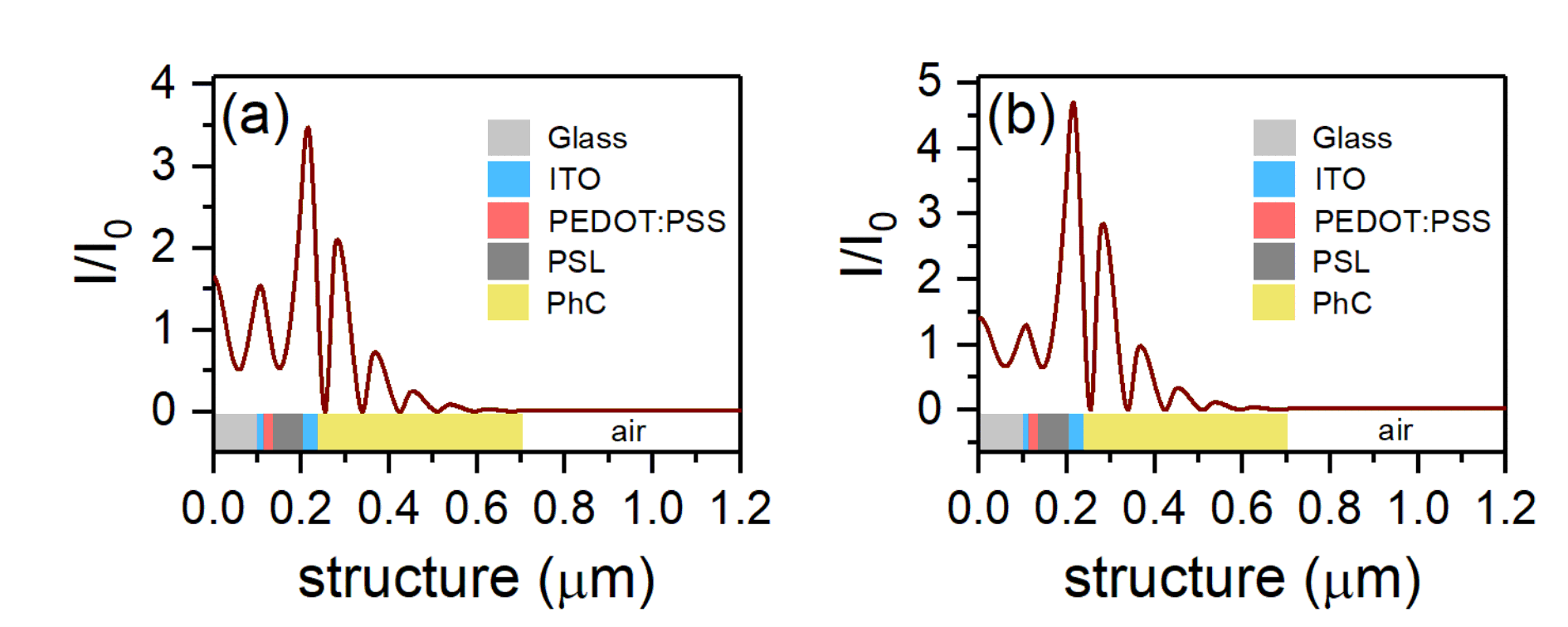}
        \caption{ Local field intensity distribution at the TPP wavelength for (a) the square and (b) hexagonal array. }
    \label{fig:field}
\end{figure}

The conjugation of the PSL containing an embedded square or hexagonal plasmonic array with the PhC will lead to the formation of a TPP at their interface; at the wavelength of the TPP, the integral absorption in the PSL, the wavelength of which is dictated by the phase matching condition, will increase. This condition can be met by changing the thickness of the ITO film adjacent to the PhC (see Fig.~\ref{fig:fig3}).     
It can be seen that, for both the square and hexagonal array, the highest integrated absorption in the PSL is obtained at an ITO film thickness of 35 nm. The increase in the integral absorption in the PSL with a decrease in the PhC period is related to the shift of the band gap to the short-wavelength spectral region, where the absorption of the (P3HT: PCBM) matrix is the highest. Thus, at the constant ITO film thickness, the highest integral absorption is obtained at a SiO$_2$ film thickness of 45~nm.
In this case, the integral absorption in the investigated wavelength range increases by $\approx$ 9.9\% for the square array and by $\approx$ 12.4\% for the hexagonal one.
In both cases, this effect is caused by the formation of a TPP localized at the interface between the PhC and the active layer doped with plasmonic nanoparticles, which is also confirmed by the spatial field distribution shown in Fig.~\ref{fig:field}. 

The field is localized at the PhC/PSL interface and decays exponentially on both sides of it.
This slight increase in the field at the TPP wavelength for both the square and hexagonal array is related to the low Q factor of the TPP~\cite{Vyunishev2019BroadbandPolariton}.

\subsection{EMT vs s-parameters}

\begin{figure}
    \centering
    \includegraphics[scale=0.5]{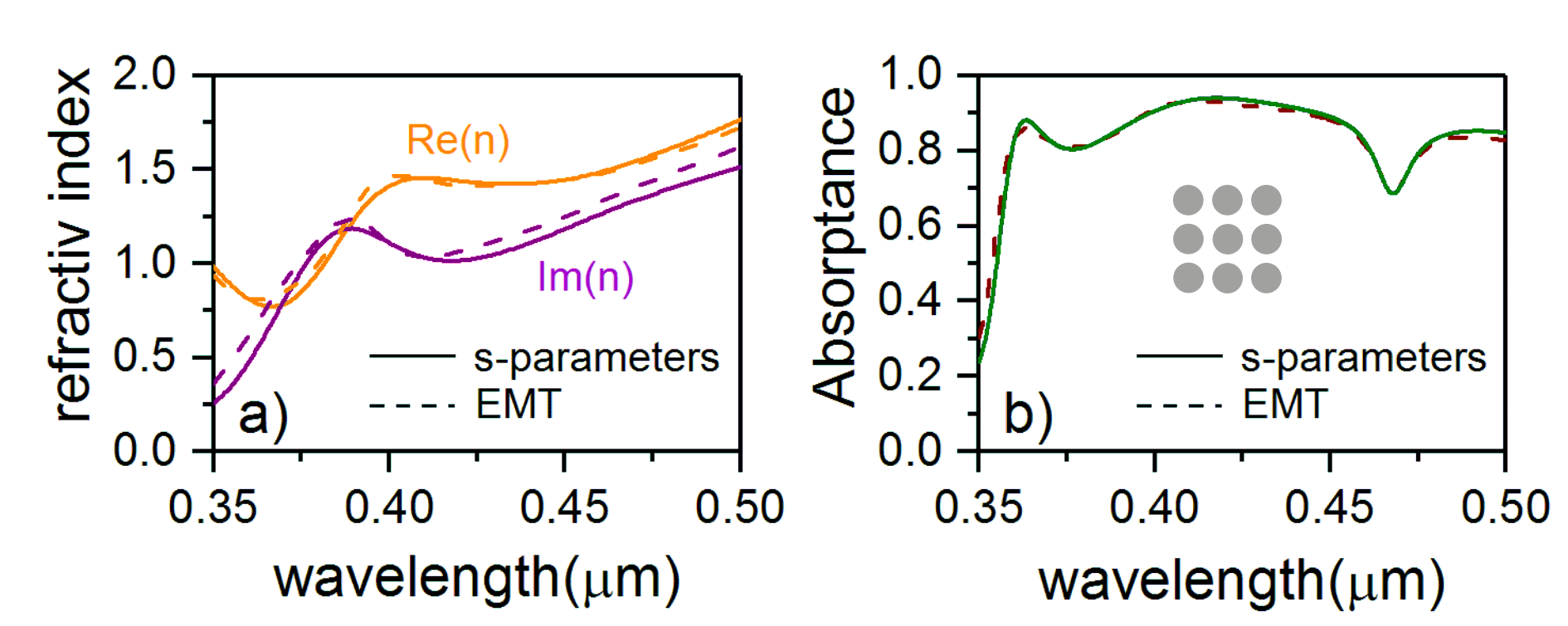}
        \caption{(a) Real and imaginary parts of the refractive index of the PSL doped with plasmonic nanoparticles. The dotted line corresponds to the calculation with the use of the effective environment model (Eq. (6)) and the solid line, to the calculation using the s-parameters for the square array. (b) Absorption in the PSL with the dispersion determined using the effective medium model and the s-parameters.}
    \label{fig:EMT}
\end{figure}

At low concentrations of silver nanoparticles in the bulk of the PSL ($0.01<f<0.3$~\cite{Spanier2000}),
the permittivity of the P3HT:PC61BM+AgNPs layer can be determined by the effective medium theory (EMT)~\cite{Maxwell-Garnett1906237}:

\begin{equation}
\varepsilon_{\text{eff}}=\varepsilon_d(\omega)\left[1+\frac{f\left(\varepsilon_m(\omega )-\varepsilon
_{d}(\omega )\right)}{\varepsilon_{d}(\omega)+\left(1-f\right)(\varepsilon_m(\omega )-\varepsilon_{d}(\omega))1/3}\right],
\label{eq:Eq1}
\end{equation}
here $f$~–~filling factor;
$\varepsilon_{d}(\omega)$ and $\varepsilon_m$($\omega $)~–~dielectric permittivity of the matrix (P3HT:PC61BM) and nanoparticles (Ag), respectively; $\omega $~–~radiation frequency.
Fig.~\ref{fig:EMT} illustrates a comparison of the dependences of the real and imaginary parts of the refractive index of the PSL doped with silver nanoparticles in a volumetric concentration of $f=20\%$ obtained using the effective environment model and the s-parameters. To obtain this bulk concentration, the array period was increased to 90 nm.

It can be seen that the dependences of the real and imaginary parts of the refractive index calculated by two different methods are in excellent agreement over the entire investigated wavelength range. Based on these data, the absorption of the OSC PSL was calculated with the determination of its dispersion using the effective medium model and the s-parameters (see Fig.~\ref{fig:EMT}b). 
It should be noted that the results obtained are in almost perfect agreement. 

Thus, the effective medium model can be used to describe the optical properties of heterogeneous media with an accuracy not inferior to that of the direct numerical calculations. In addition, the use of such models can significantly speed up the calculation and optimization of optical structures.

\section{Conclusion}

The spectral properties of a model of an organic solar cell based on a Tamm plasmon polariton localized at the interface between a photonic crystal and a photosensitive layer with an embedded square or hexagonal plasmon array were studied.
The effective refractive index of the photosensitive layer was determined by the s-parameter retrieval method; the energy spectra of the structure and its local intensity distribution were calculated by the transfer matrix method.
In the proposed model it was shown that the integral absorption in the photosensitive layer with the introduced square and hexagonal arrays increases by 10\% and 12\%, respectively, as compared with the solar element without the Tamm plasmon polariton.
The comparative computation of the integral absorption in the photosensitive layer with the optical properties determined using the effective medium model and the direct numerical calculation was made. It was shown that, at low (about 20\%) concentrations of nanoparticles in the bulk of the photosensitive layer, the results obtained by two different methods agree well. Thus, the Maxell--Garnett model can be used at the first (estimation) stages of the calculation of the optical properties of such structures with an accuracy no lower than that of the direct calculation.

\section*{Funding}
The reported study was funded by the grant of the President of Russian Federation $N\textsuperscript{\underline{o}}$ MK-46.2021.1.2 and by Russian Foundation for Basic Research, Government of Krasnoyarsk Territory,
Krasnoyarsk Region Science and Technology Support Fund
to the research project $N\textsuperscript{\underline{o}}$ 19-42-240004.

\bibliography{mendeley,references}

\end{document}